# Electrical activation and electron spin coherence of ultra-low dose antimony implants in silicon

T. Schenkel[1,*], A. M. Tyryshkin[2], R. de Sousa[3], K. B. Whaley[3], J. Bokor[1, 4], J. A. Liddle[1], A. Persaud[1], J. Shangkuan[1,3], I. Chakarov[5], and S. A. Lyon[2]

[1]E. O. Lawrence Berkeley National Laboratory, Berkeley, CA 94720

[2] Department of Electrical Engineering, Princeton University, Princeton, NJ 08544

[3]Department of Chemistry and Pitzer Center for Theoretical Chemistry, University of California, Berkeley, CA 94720

[4]Department of Electrical Engineering and Computer Science, University of California, Berkeley, Ca 94720

[5]Silvaco International, 4701 Patrick Henry Dr., Santa Clara, CA 95054

We implanted ultra low doses ($2\times10^{11}$ $cm^{-2}$) of $^{121}Sb$ ions into isotopically enriched $^{28}Si$ and find high degrees of electrical activation and low levels of dopant diffusion after rapid thermal annealing. Pulsed Electron Spin Resonance shows that spin echo decay is sensitive to the dopant depths, and the interface quality. At 5.2 K, a spin decoherence time, $T_2$, of 0.3 ms is found for profiles peaking 50 nm below a $Si/SiO_2$ interface, increasing to 0.75 ms when the surface is passivated with hydrogen. These measurements provide benchmark data for the development of devices in which quantum information is encoded in donor electron spins.

*Email: T_Schenkel@LBL.gov

Spins of electrons bound to donor atoms in silicon at low temperature are promising candidates for the development of quantum information processing devices [1-3]. This is due to their long decoherence times, and the potential to leverage fabrication finesse in a silicon transistor paradigm. Recently, relatively long transverse relaxation times ($T_2$) were determined for electron spins in pulsed electron spin resonance (ESR) studies of phosphorous donors in isotopically enriched silicon. Here, donors were present as a random background doping across $^{28}$Si epi layers and $T_2$ extrapolated to 60 ms for isolated donors [3]. Formation of test devices for quantum information processing requires the integration of individual dopant atoms with a control and readout infrastructure. Donor array fabrication is being addressed by ion implantation [4-6] and scanning probe based hydrogen lithography [7, 8]. Dopant spacing depends on the choice of entangling interactions between quantum bits (qubits) and ranges from 20 to over 100 nm, corresponding to ultra low ion implantation doses of $<10^{10}$ to $2.5\times10^{11}$ cm$^{-2}$. In this letter, we report on depth profiles and electrical activation following rapid thermal annealing (RTA) of ultra low dose $^{121}$Sb implants and correlate electron spin relaxation times with the dopant distribution below an interface and with the interface quality.

We processed wafers with 10 μm thick, $^{28}$Si enriched epi layers (500 ppm $^{29}$Si) on p-type natural silicon (100) and natural silicon control wafers (100), both with impurity concentrations $\leq10^{14}$ cm$^{-3}$. Standard CMOS processes were followed for formation of 5-10 nm thick thermal $SiO_2$. Typical densities of trapped charges and interface traps were 1 to $2\times10^{11}$ cm$^{-2}$ for the thermal oxides. $^{121}$Sb-ion implantation with a dose of $2\times10^{11}$ cm$^{-2}$ was conducted with implant energies of 120 keV and 400 keV. $^{121}$Sb was used to avoid any ambiguity of results due to $^{31}$P background in $^{28}$Si epi layers. RTA for repair of implant damage and substitutional incorporation of dopants into the silicon lattice, i. e., electrical activation, was performed with an AGA Heatpulse 610. Following annealing, carrier depth profiles were probed with Spreading Resistance Analysis (SRA) [9]. Secondary Ion Mass Spectrometry (SIMS) [9] was used to characterize elemental depth profiles in as-implanted and annealed samples. As-implanted depth profiles were also simulated using a dynamic Monte Carlo model [10]. Electron spin relaxation in $^{28}$Si samples was probed by pulsed ESR in an X-band (9.7 GHz) Bruker EPR spectrometer at temperatures of 5 to 10° K. Standard 2-pulse echo and inversion recovery experiments were used to measure $T_1$ and $T_2$ relaxation times, respectively [3].

Figure 1 a) shows SIMS and SRA depth profiles of an $^{28}$Si sample implanted with $^{121}$Sb at 400 keV (0° tilt) and a dose of $2\times10^{11}$ cm$^{-2}$, together with a simulation of the as-implanted profile. The sample was annealed in an $N_2/H_2$ ambient at 980° C for 7 s. We define the electrical activation ratio as the ratio between the carriers from an integrated, background corrected SRA profile, and the implanted dose. Nominal implant doses agreed with values extracted from SIMS spectra. The electrical activation of donors in the $^{28}$Si sample is complete, i. e., 100%, within the accuracy of the measurements.

Due to low donor concentrations of $10^{16}$ cm$^{-3}$ and below, the SIMS spectra are relatively noisy and yet allow some definite conclusions. The as-implanted SIMS spectrum matches the simulated depth profile closely, but the SIMS spectrum of the annealed $^{28}$Si sample shows significant broadening compared to the as-implanted profile. This broadening is reproduced in the SRA profile. We attribute the profile broadening to diffusion during RTA. The profile broadening is symmetrical and there is no evidence for segregation of dopants towards the $Si/SiO_2$ interface. This is important and in contrast to recent observations of phosphorus segregation towards $Si/SiO_2$ interfaces during RTA [5]. Antimony diffuses through a vacancy mechanism and diffusion is retarded by the interstitials injected from the $Si/SiO_2$ interface during RTA [11, 12].

In Figure 1 b), we show SIMS, SRA and simulated depth profiles from samples implanted with $^{121}$Sb ($2\times10^{11}$ cm$^{-2}$) at 120 keV (7° tilt). The sample was annealed in an $N_2$ ambient at 1000° C for 10 s. Simulations of the as-implanted profile and the SIMS profile of the annealed sample agree very well. SIMS spectra of the as-implanted samples were not available. The SRA spectrum shows a carrier distribution that is strongly shifted towards the surface, and the apparent electrical activation is only 3%. Repeated SRA measurements on these samples showed inconsistent results. Apparent low carrier concentrations in SRA were consistent with ESR measurements. Signal in both SRA and ESR stems from donors that are incorporated on substitutional sites in the silicon lattice and that are electrically neutral. Band bending at the interface due to interface and oxide charges leads to the ionization of donors close to the interface and thus to the reduced number of active donors seen by SRA and ESR. In addition, we observed that illumination of samples at low temperatures during ESR measurements increases the spin counts significantly for samples with apparent low electrical activation levels in SRA. The close agreement between the simulated as-implanted, and the SIMS profile of the annealed sample indicates only minimal diffusion and no segregation of donors to the interface.

Electron spin relaxation times were measured for shallower (120 keV implant energy, peak dopant depth of 50 nm) and deeper (400 keV implant energy, peak dopant depth of 150 nm) $^{121}$Sb implants in $^{28}$Si with thermal $SiO_2$ interface. The insert of Figure 2 shows the six-line ESR spectrum of implanted $^{121}$Sb donors with splitting arising from hyperfine interaction with the nuclear spin (I=5/2) of $^{121}$Sb. Most of the relaxation measurements were done at the M=1/2 hyperfine line but the relaxation times were identical on other lines. For shallower donors, we found spin relaxation times $T_1$ = 15 ± 2 ms and $T_2$ = 0.3 ± 0.03 ms at 5.2° K. For deeper implants, a much longer $T_2$ = 1.5 ± 0.1 ms was measured while the spin-lattice relaxation time, $T_1$ = 16 ± 1 ms, did not change.

The thermal oxide layer of $^{28}$Si samples was then removed by etching in a hydrofluoric acid solution, resulting in a hydrogen terminated silicon (100) surface of modest quality (compared to Si-111) [13, 14]. The electrical properties of the H-Si interface were not probed here, but typical interface trap densities below $10^9$ cm$^{-2}$ have previously been reported [15]. Following hydrogen passivation, $T_2$ increased to 2.1 ms for the 400 keV implants, and to 0.75 ms for the 120 keV implants. We summarize our results in Table 1.

Dopants diffuse through interaction with interstitials and vacancies. In the ultra low dose implant regime, dopant redistribution during RTA is affected by defect injection from the dielectric-silicon interface, and by the interaction of dopants with point defects that did not recombine following the slow down of implanted ions (i. e., transient enhanced diffusion) [11, 12, 16]. For the heavier antimony ions, point defect formation is enhanced in collision cascades compared to implantation of boron or phosphorus. Recombination of vacancies and interstitials is incomplete for ultra low dose implants [16]. From SIMS measurements and simulations we find that dopant redistribution is minimal for the 120 keV implants. The enhanced diffusion for the 400 keV implants might be a result of less complete recombination of vacancies and interstitials in more extended collision cascades formed by higher energy ions.

Quantum computer test structures require both efficient electrical activation of dopants and retention of initial dopant positions during thermal processing. The results shown here fulfill these requirements, and enable testing of single spin readout architectures with high device yields. Further optimization of thermal processing with defect engineering and control of interface properties promises to allow high levels of substitutional incorporation and minimal dopant redistribution also for shallow implants (with depth of ~20 to 30 nm below the surface). Probing of spin dynamics in shallow donor implants requires gate control over energy levels to avoid donor ionization due to band bending, e. g. by tuning into a flat-band condition.

Donor electron spin relaxation is correlated both with the depth distribution of dopants with respect to an interface and with the interface quality (Table 1). The fact that removal of the oxide layer and H-passivation of the interface lead to a significant increase in $T_2$ allows us to conclude that it is coupling to paramagnetic defects [17] in the oxide and at $Si/SiO_2$ interface which limits $T_2$ for both shallow and deeper dopant distributions. A likely mechanism is fluctuating magnetic fields due to spin flips of paramagnetic defects and the loading and unloading of traps at the interface and in the oxide. At much reduced interface trap levels for the H-Si surface, coherence is likely limited by instantaneous diffusion, i.e. magnetic dipole coupling of neighboring dopant atoms [2, 3], and possibly other effects that have not been quantified. The effect of nuclear spins of hydrogen atoms at a coverage of $6.8\times10^{14}$ cm$^{-2}$ for an H-Si (100) surface on donor electron spin coherence can be estimated with a spectral diffusion model [2]. This process is expected to limit $T_2$ only for times longer than 1 s for the profiles shown in Figure 1 and is probably not important in the present case, but will become important for shallower implants.

In summary, annealing of ultra low dose antimony implants in isotopically enriched $^{28}Si$ leads to high degrees of electrical activation with minimal diffusion. The transverse electron spin relaxation time, $T_2$, increases when dopants are placed deeper below a thermal $SiO_2$ interface, and hydrogen passivation of the silicon surface yields an even longer $T_2$ of 2.1 ms (at 5.2 K), indicating that spin flips in paramagnetic defects limit coherence in the presence of a $Si/SiO_2$ interface. Spin coherence times well in excess of 1 ms are readily achieved with standard silicon processing, enabling tests of quantum information processing architectures with donor electron spin qubits.

**Acknowledgments**

We thank T. C. Shen for stimulating discussions, and the staff of the UC Berkeley Microlab for their technical support. This work was supported by NSA under ARO contract number MOD707501, the U. S. DOE under contract No. DE-AC02-05CH11231, and by NSF under Grant No. 0404208. Work at Princeton was supported by ARO and ARDA under Contract No. DAAD19-02-1-0040. KBW and RdS acknowledge support by the DARPA SPINs program under grant No. FDN0014-01-1-0826.

**Figure and table captions:**

Figure 1: a) SIMS and SRA profiles of as-implanted and annealed samples together with simulated depth profiles for antimony ($^{121}$Sb) implanted to a dose of $2\times10^{11}$ cm$^{-2}$ with an implant energy of a) 400 keV $^{121}$Sb and b) 120 keV.

Figure 2.: The 2-pulse electron spin echo (ESE) decay for $^{121}$Sb donors in $^{28}$Si (400 keV, dose $2\times10^{11}$ cm$^{-2}$, annealed) measured at the M = +1/2 line in the ESR spectrum at 5.2° K. Because the ESE signal decay is strongly suppressed at long $\tau > 0.5$ ms by magnetic field noise [3], the exponential fit (dashed line) was calculated using only $\tau < 0.5$ ms and resulted in $T_2 = 2.1$ ms. The insert shows an ESR spectrum of $^{121}$Sb, consisting of six narrow (<0.2 G) lines split by hyperfine interaction with the nuclear spin (I=5/2) of $^{121}$Sb. The broad feature in the centre of the spectrum is from surface defects due to sample preparation.

Table 1: Summary of activation ratios and decoherence times for oxide and hydrogen passivation of $^{28}$Si surfaces (100).

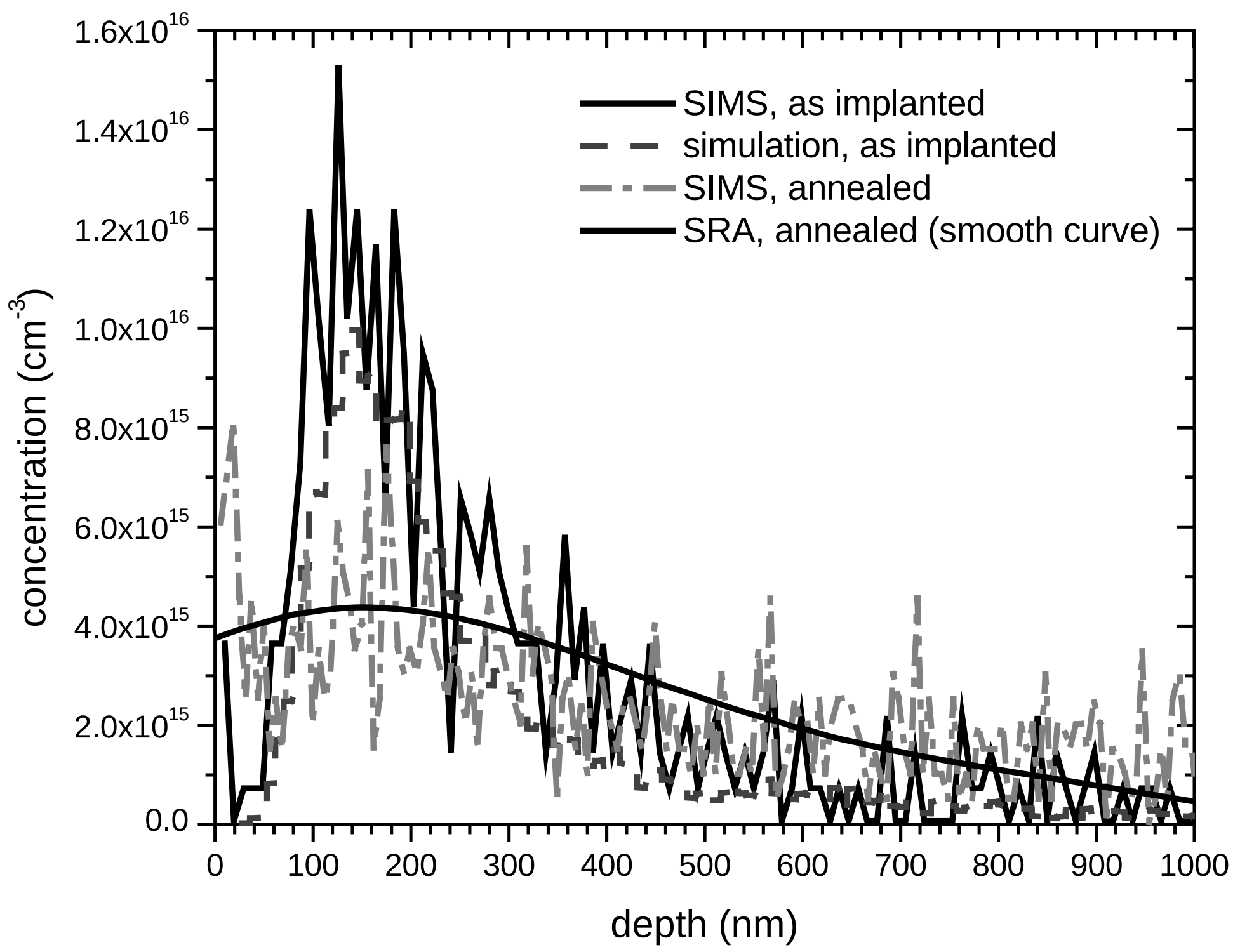

Figure 1 a)

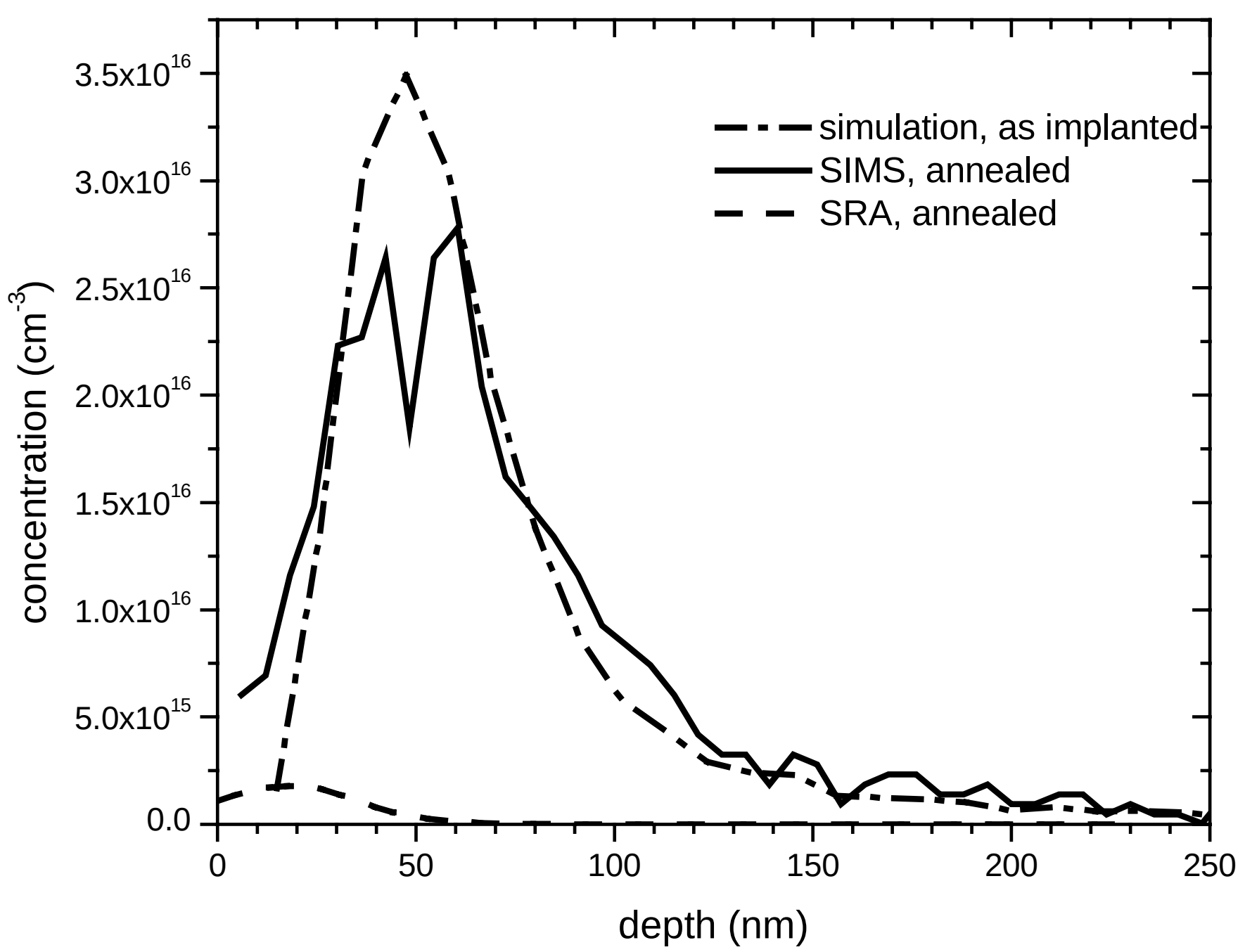

Figure 1 b)

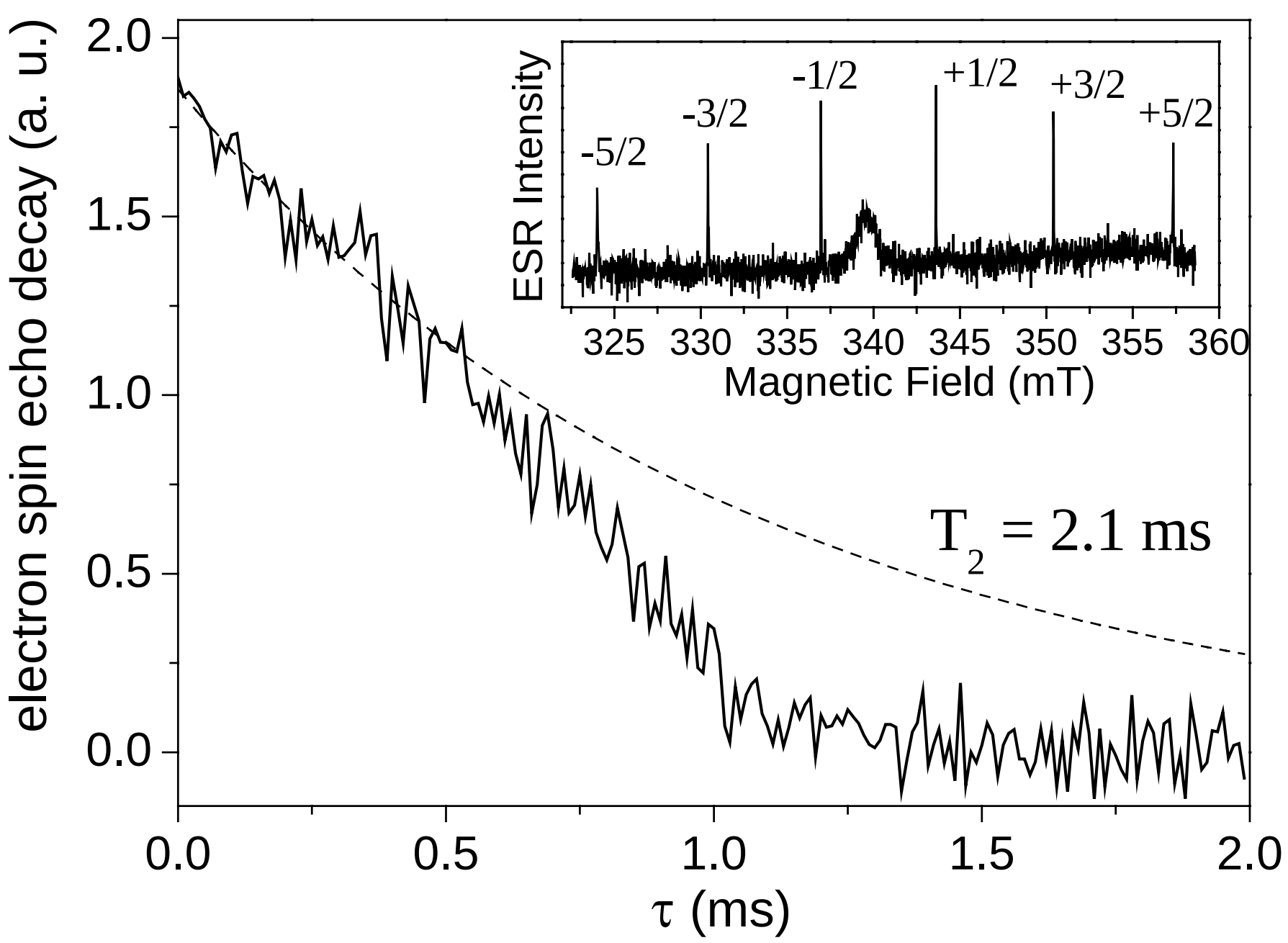
2.0
1.5
1.0
0.5
0.0
electron spin echo decay (a. u.)
0.0
0.5
1.0
1.5
2.0
τ (ms)
T2 = 2.1 ms
ESR Intensity
-5/2
-3/2
-1/2
+1/2
+3/2
+5/2
325 330 335 340 345 350 355 360
Magnetic Field (mT)

Figure 2

| Interface | Peak dopant depth (nm) | Apparent activation ratio | $T_1$ (ms) at 5.2° K | $T_2$ (ms) at 5.2° K |
|---|---|---|---|---|
| $SiO_2$ | 50 | 3.4 % | 15 ± 2 | 0.30 ± 0.03 |
| H-Si | 50 | - | 16 ± 2 | 0.75 ± 0.04 |
| $SiO_2$ | 150 | 100 % | 16 ± 1 | 1.5 ± 0.1 |
| H-Si | 150 | - | 14 ± 1 | 2.1 ± 0.1 |

Table 1